\documentclass[fleqn,10pt]{wlscirep}

\usepackage{graphicx}%
\usepackage{multirow}%
\usepackage{amsmath,amssymb,amsfonts}%
\usepackage{amsthm}%
\usepackage{mathrsfs}%
\usepackage[title]{appendix}%
\usepackage{xcolor}%
\usepackage{textcomp}%
\usepackage{manyfoot}%
\usepackage{booktabs}%
\usepackage{algorithm}%
\usepackage{algorithmicx}%
\usepackage{algpseudocode}%
\usepackage{listings}%
\usepackage{float}

\title{Using large scale GPS data to reveal EV driver activity patterns beyond charging sessions}

\author[1,2,3]{Callie Clark}
\author[1]{Anne Driscoll}
\author[2,4]{Xiyuan Ren}
\author[2,4]{Salsabil Salah}
\author[5]{Marta C. Gonz\'alez}
\author[2,4]{Joseph Y. J. Chow}
\author[1,6,*]{Takahiro Yabe}
\affil[1]{Center for Urban Science \& Progress, Tandon School of Engineering, New York University, NY 11201, USA}
\affil[2]{Department of Civil, Urban, and Environmental Engineering, Tandon School of Engineering, New York University, NY 11201, USA}
\affil[3]{Marron Institute, New York University, NY 11201, USA}
\affil[4]{C2SMART Center, Tandon School of Engineering, New York University, NY 11201, USA}
\affil[5]{Department of Civil and Environmental Engineering, University of California at Berkeley, CA, USA}
\affil[6]{Department of Technology Management and Innovation, Tandon School of Engineering, New York University, NY 11201, USA}

\affil[*]{takahiroyabe@nyu.edu}

\begin{abstract}
Accurate insights into electric vehicle (EV) driver behavior are essential for long-term infrastructure planning, grid management, and understanding downstream economic impacts, yet individual level data on EV mobility remains limited. 
Here, we develop a scalable framework to infer EV ownership and charging behavior from passively collected, high-resolution mobility traces covering over 760,000 drivers across four major U.S. metropolitan areas. 
We identify likely EV drivers based on distinctive visitation patterns to charging stations and gas stations, frequency of visits, and daily travel behavior, and calibrate cohort size using aggregate EV registration statistics. 
The resulting EV cohort closely matches official registration data at the zip code level (Pearson correlation of 0.73) and exhibits charging patterns consistent with independent, charger level benchmark datasets, providing external validation of the inferred population. 
Leveraging this inferred cohort, we reconstruct charging events and associated activity patterns to examine how EV drivers interact with surrounding urban amenities. 
Compared to non-EV drivers, EV drivers exhibit systematically higher visitation rates to nearby cafes and restaurants during charging sessions, revealing significant economic spillover effects. Furthermore, we find EV drivers exhibit trip bundling behavior, visiting more POIs over less time and distance on days where they charge versus all other days. These patterns are not observable in conventional charging session data, which lack behavioral context beyond the charging event itself.
Our results demonstrate the potential of using mobility data to enable a richer, behaviorally grounded understanding of the ``off-plug'' needs of EV drivers, providing a foundation for optimizing charging infrastructure deployment and co-locating complementary urban amenities in an increasingly electrified transportation landscape.
\end{abstract}
\begin{document}

\flushbottom
\maketitle
%
%
\thispagestyle{empty}







\maketitle

\section*{Introduction}\label{sec1} 




Electric vehicle charging stations (EVCS) are an increasingly important component of transportation infrastructure \cite{brown2024electric}. Electric vehicles account for roughly 10\% of new vehicle sales in the United States (US) and 20\% globally in 2024 \cite{IEA2025GlobalEV}, and in the US, substantial federal investment is accelerating their deployment. As battery technology advances and charging networks densify, EV uptake is expected to accelerate the transition toward electrified transport \cite{BNEF2025EVOutlook}. 
As EVs move beyond early adopters, who were disproportionately concentrated among higher-income households, EV driver behavior is becoming more heterogeneous and less predictable, challenging assumptions derived from conventional vehicle use and earlier stages of adoption. 
Moreover, recent work suggests that EV adoption itself is shaped by spatial and social network effects among potential adopters, which can substantially alter long-term forecasts~\cite{wu2025planning}. As adoption scales, these dynamics amplify the economic and grid level consequences of EV driver behavior, raising important policy considerations for equitable infrastructure and energy planning \cite{lou2024income,yu2025equity}. 

Historically, research in this domain has relied on surveys \cite{lee2020exploring,tal2020factors, bhat2024preferences, visaria2022user}, field studies \cite{franke2013understanding, bjornsson2015plugin, sun2016fast, yu2016modeling}, and station level session data \cite{bauer2021charging, zheng2024effects, huang2024unveiling, kim2017heterogeneous}. While surveys and field work provide high-resolution insights into driver intent, they are inherently limited by high costs, small sample sizes, and poor longitudinal scalability. Conversely, station level data offers the scale necessary for urban analysis but suffers from contextual blindness, as it captures only the duration and energy of the plug-in event while remaining unaware of the driver’s movements before and after the session. Consequently, it remains challenging to understand EV driver behavior that are not directly mediated by the charging event, including visitation to nearby amenities before, during, and after charging, and how drivers integrate charging into their daily activity patterns.

The increase in availability and quality of mobility data in the past several decades~\cite{gonzalez2008understanding} has enabled other fields to address similar data constraints, including transportation mode choice~\cite{huang2019transport}, activity patterns\cite{ren2022random, fu2025activity}, and trip purpose \cite{alsger2018public}. In the early stages of EV adoption, EV behavior and EV owner characteristics were more distinct and predictable due to the high cost of EVs and the more limited battery range. As a result, income and vehicles miles traveled (VMT) were sufficient to probabilistically estimate EV ownership and study their mobility and charging behavior \cite{xu2018planning}. However, as EV adoption expands and technology evolves, these signals alone are becoming less informative, while the growing scale of mobility data creates new opportunities to study EV behavior more directly. Recent studies have leveraged the staggered deployment of new EVCS with nearby points-of-interest (POI) visitation data to demonstrate economic spillovers to nearby businesses, finding that installations can significantly increase annual consumer spending at nearby businesses \cite{zheng2024effects,de2025causal,Babar2024}. However, these findings are limited to aggregate trends, masking the individual level behavioral mechanisms that drive these patterns. 

To bridge these gaps, we propose a novel framework using individual level mobility traces to identify and validate a cohort of EV drivers across four different U.S. cities. Using anonymized, high-granularity device location data from Cuebiq \cite{cuebiq_data}, we move beyond the constraints of infrastructure and survey data to analyze real-world EV mobility at an unprecedented scale. Our proposed model uses longitudinal features to distinguish EV drivers from internal combustion engine (ICE) drivers. We find that mobility-related variables, including the number of monthly stops at gas stations, the number of unique EVCS visited, having more EVCS visits than gas station visits, the average daily vehicle miles traveled (VMT), and such sociodemographic characteristics as the composition of housing types in the neighborhood, are significant factors that predict individual level EV ownership. Our model is validated by comparing with session level charging data provided by EV Watts \cite{livewire2025}, and we also verify that EV driver behavior is significantly different from the general population. 
We further show that the proposed model can robustly identify EV drivers across four diverse American metropolitan areas: the San Francisco Bay Area, Seattle, Boston, and Denver. Lastly, using our inferred EV driver cohort, we explore previously unobservable EV driver behavior and find that EV drivers have distinct POI visitation patterns during a charging session, demonstrating a higher propensity to visit restaurants compared to all other visitors in the same built environment. We also find significant differences in mobility behavior among EV drivers on days with a public charging session, indicating that these behavioral changes extend throughout the entire day.

This approach provides a scalable lens into modern EVCS public charging, offering insights into how these drivers navigate the city both while tethered to a charger and during their broader daily routines. By capturing granular movements, we can identify behaviors such as trip-chaining patterns and mobility outside of charging sessions that underpin larger economic shifts. We further find that the proposed model generalizes well across multiple cities, enabling its application to new geographic locations without requiring additional validation data, which are often difficult to obtain. These capabilities offer a foundation for more behaviorally informed urban planning, infrastructure placement optimization, and energy system management in rapidly electrifying cities.

\section*{Results}\label{sec2} 








\subsection*{Challenges in detecting EV drivers from mobility traces}

Using high-resolution GPS mobility data spanning four major U.S. metropolitan statistical areas (MSAs) over a three month period in 2022, we observe daily individual mobility patterns for over 760,000 anonymized and opted-in individuals. 
We use an anonymized location dataset of mobile phones and smartphone devices provided by Cuebiq~\cite{cuebiq_data}, a location data intelligence company that collects anonymous, privacy compliant location data of mobile devices. Cuebiq processes data collected from mobile devices whose owners have actively opted in to share their location, resulting in roughly 15 million daily active users in the United States. This detailed longitudinal dataset includes both device-level timestamped GPS pings and algorithmically inferred ``stops'', enabling fine-grained reconstruction of individual mobility trajectories. 
To contextualize mobility behavior, we integrate these individual trajectories with point-of-interest (POI) location data including geolocated electric vehicle charging stations (EVCS) and gas stations compiled from the Alternative Fuel Data Center (AFDC)~\cite{afdc_ev_locations}, Open Charge~\cite{openchargemap_2025} and Safegraph POI databases~\cite{safegraph_global_places_2022}, as illustrated in Figure 1a. 

We identify fueling-related events directly from mobility traces. Because EV drivers may plug in and leave immediately from the charging station, which could result in no ``stop'' being recorded, we rely on device-level GPS pings to detect potential charging sessions. We define a potential EVCS charging session as an instance where we have the following conditions: 1) at least two observations within 10 meters of an EVCS, 2) separated by at least 10 minutes, 3) with at least one observation recording a speed below 10 km/h, and 4) with intermediate movement confined within one kilometer of the station and below 20 km/h. In essence, we ensure that the driver slows down near a charging point, returns to a charging point, and does not move in uncharacteristic ways (e.g., too far away from the charging station or at a high speed) during charging. In contrast, gas stations occupy larger footprints and require drivers to remain  near their vehicles; we therefore use the provided ``stops'' dataset, defining a gas station visit as a stop within 15 meters of the gas station and a stay duration between 2 to 15 minutes (see Methods). Applying this procedure to all users' individual mobility trajectories across three months, we identify 127,730 potential EVCS station visits and 1,382,679 gas station visits for 761,666 users. As shown in Figure 1b and 1c, while around 42\% of users have 1 or more stops near gas stations, only around 6.3\% of the users have 1 or more potential EV charging sessions.  

A natural starting point is to label users with at least one detected potential EV charging session as EV drivers. However, this approach substantially overestimates EV ownership: the projected EV ownership rate derived from this rule in the San Francisco area is 9.2\%, which is a significant overestimation of EV ownership, compared to the ground truth of 4.4\% \cite{cadmv2019}. This discrepancy arises because EV charging stations are often located within parking lots, where conventional parking spaces neighbor the charging port. Given the uncertainty in GPS positions, a true EV charging event isn't distinguishable from a vehicle parked nearby. One way to mitigate this issue is to analyze longitudinal behavior patterns and identify consistent and repeated visits to charging stations. 

Misclassifications of charging events are further amplified by the heterogeneity of charging behavior. EV drivers rely on a mix of home, workplace, and public charging, with 18.4\% of EV drivers never using a DC fast charger or supercharger and 26.9\% of them never using level 2 public chargers \cite{turner2025ev}.
Because residential and workplace locations are privacy-obscured in the mobility dataset and private EVCS location data are sparse, a substantial portion of EV charging that occurs at home or a workplace is unobservable at the individual level.

This creates a systematic sampling bias: reliance on public charging station visits disproportionately capture drivers without private infrastructure, such as residents of multifamily housing or long-distance travelers, while under representing higher income owners with home garages. 
As a result, EV drivers exhibit highly variable visitation patterns across time and space, making it challenging to infer EV ownership from any single behavioral signal. 

Together, these factors render simple threshold-based classification of charging sessions insufficient. EV driver identification is therefore not a single-event detection problem, but must be inferred from longitudinal fueling patterns that distinguish consistent charging behavior from incidental parking near chargers. By modeling repeated interactions with both EVCS and gas stations over time, we move beyond proximity-based labeling toward a behavioral framework that robustly separates electric from internal combustion vehicle users.




\begin{figure}
    \centering
    \includegraphics[width=1\linewidth]{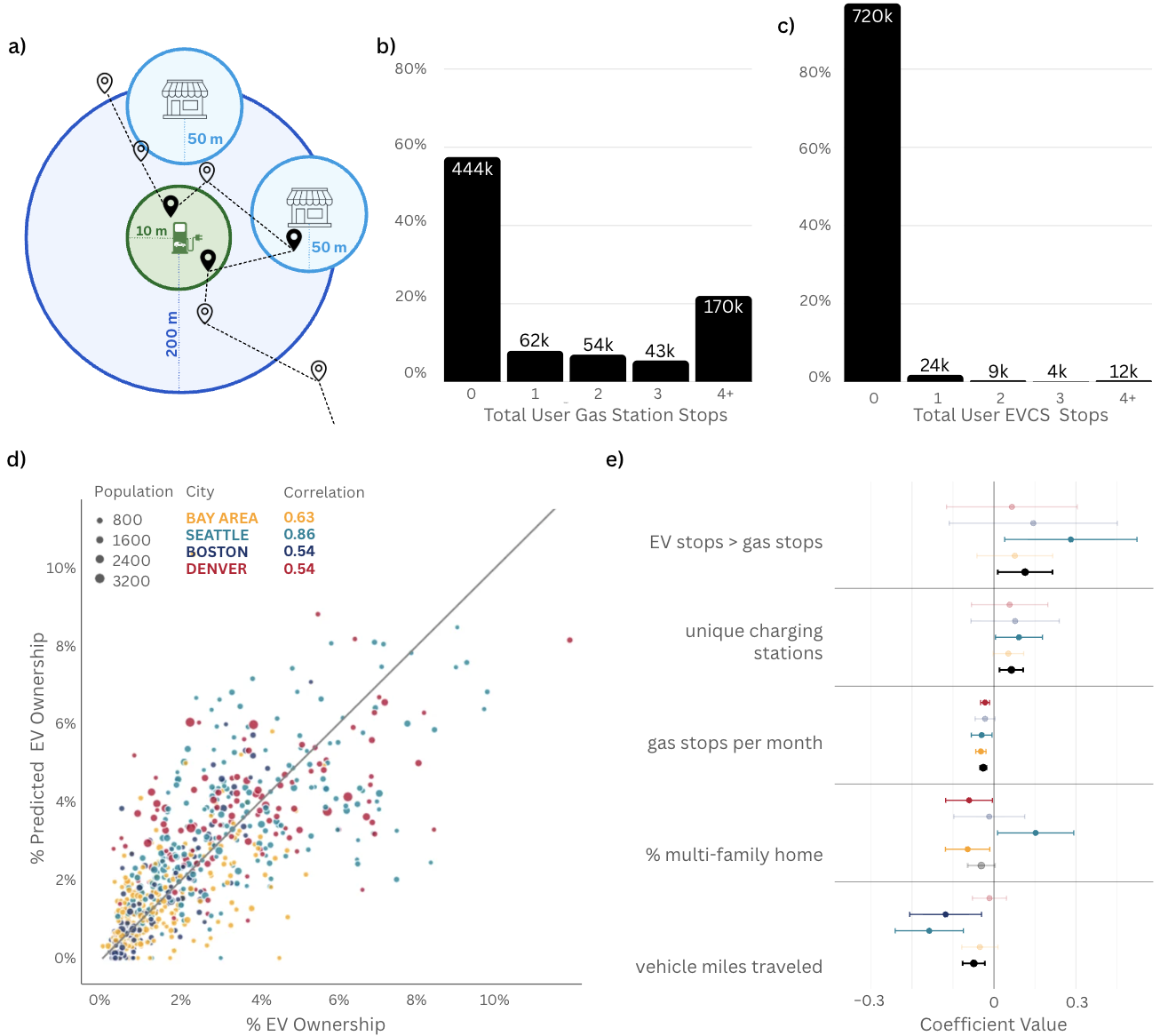}
    \caption{\textbf{EV cohort is identified from mobility data using a linear model.} \textbf{a)} Charging sessions are inferred by spatially intersecting device-level GPS pings with EVCS geofences, while gas station and POI visits are identified using pre-constructed GPS ``stops''. \textbf{b)} Frequency of detected gas station stops and \textbf{c)} EV charging sessions across the study population, highlighting the rarity of public charging events relative to conventional fueling. \textbf{d)} Scatter plot of predicted versus actual EV ownership rates at the zip code level across the four study MSAs. The zip level Pearson correlation is 0.73 for the pooled model.  \textbf{e)} Regression coefficients illustrate the behavioral signatures used to rank the EV cohort. Coefficients for the pooled model are shown (black), as well as coefficients for the city-specific models. A higher ratio between EV stops and gas station stops, the number of unique EVCS visited, and vehicle miles traveled are the strongest predictors of EV ownership.}
    \label{fig:fig1}
\end{figure}

\subsection*{Inferring EV ownership from longitudinal mobility patterns}

To move beyond identifying EV drivers based solely on the presence of a charging event, we constructed a feature set that captures longitudinal mobility patterns over a three month period in spring of 2022. In addition to the counts of EVCS and gas station visits, we incorporate the unique number of stations visited, the monthly visitation frequency, average daily vehicle miles traveled, and demographic information inferred from the users' home census block groups. We then trained a pooled linear model across all four cities to estimate EV ownership at the individual level. By pooling data across four diverse metropolitan areas, the model leverages a broader variance in urban topology and charging infrastructure density, ensuring that the identified behavioral fingerprints are not idiosyncratic to a single city.

Because individual level ownership labels are non-existent due to privacy protections, prediction must rely on aggregate registration data, which are reported at the zip code level. This introduces an ecological inference challenge: the risk that the model might capture neighborhood level socioeconomic proxies rather than true individual behavior. EV ownership is strongly correlated with demographic characteristics such as income and race, therefore incorporating coarse geographic variables directly as individual predictors can  dominate the model, effectively reducing it to a proxy for neighborhood affluence. To mitigate this, we prioritize features derived from individual longitudinal traces, such as the ratio of EVCS to gas station visits, while incorporating city level fixed effects to account for baseline differences in regional adoption rates and infrastructure maturity. As a result, we limit the use of demographic variables, with the exception of multi-family housing rates, which capture structural constraints on access to private charging infrastructure. This structure allows the model to learn the specific feature weights that characterize high-adoption areas, which can then be applied back to individual users to generate an ownership probability score. 

We then use the individual level ownership probabilities to calibrate the EV driver cohort size using known metropolitan EV ownership information. Specifically, users are ranked by predicted propensity of being an EV driver, and the cohort is selected to match the reported zip code level EV ownership rate of 4.4\% in the Bay Area, 1.9\% in Seattle, 1.8\%in Boston, and 5.5\% in Denver. Then we adjusted downward by 26.9\% to account for EV drivers who report never using level 2 public charging infrastructure~\cite{turner2025ev}, resulting in 19,228 EV drivers across the four cities. The inferred EV cohort is then aggregated to the zip code level and compared against the original training data of zip level EV registration rates. 



The resulting model identifies a cohort that mirrors official EV registration data with high fidelity, achieving a zip level Pearson correlation of 0.73 across all cities, as shown in Figure 1d. We benchmark this model against two baselines: a heuristic model where we define an EV driver as any user with one or more EV sessions and no gas station visits in the entire study period, and a baseline model that uses VMT and median income to predict EV ownership. Our model has significantly higher correlation compared to the heuristic model (Pearson correlation of 0.40), and the baseline model (Pearson correlation of 0.55) (see Supplemental for full regression coefficients and full validation analysis). This performance improvement underscores the importance of incorporating longitudinal visitation patterns and spatial context of fueling behavior in identifying EV drivers. The final model is a pooled regression across all four cities, incorporating city level fixed effects.  Spatial distribution of inferred EV owners and state-reported registration numbers across four major MSAs are shown in Figure 2a-2d.

Regression coefficients shown in Figure 1e further reinforce this behavioral interpretation (see Supplementary Table S1 for the full regression coefficient table). The strongest predictors of EV ownership are having more EVCS sessions than gas station visits, and more unique charging stations visited. Notably, the prevalence of multi-family housing (MFH) is statistically insignificant once mobility features are included, suggesting that mobility data-based longitudinal behavioral signals are sufficient to identify EV drivers even in dense urban environments.
While coefficient estimates vary across cities, they are directionally consistent with expectations. The MFH variable remains ambiguous: although EV owners are more likely to live in single-family homes overall, those who rely on public charging are disproportionately MFH residents \cite{hsu2021public, kuby2025ev}.

\subsection*{Inferred EV cohort validates against nationwide charging session data} 

To evaluate the inferred EV cohort, we benchmark their behavior against EV Watts data, an independent database of EV charging session data, and available survey information. The EV Watts database provides session level charging information at a large scale, with data coverage for the Bay Area, Seattle, Boston and Denver totaling over 127,000 charging sessions over our three month study period (spring of 2022). From these observed sessions, we can construct city-specific, session-level information like duration of charging and temporal charging patterns to use as a benchmark. Additionally, we compare our EV cohort against the entire user sample to evaluate if our EV cohort is capturing distinct behavior.

Our first benchmark to evaluate the success of our EV cohort is comparing  vehicle miles traveled (VMT) between our EV cohort and the remaining users (presumed ICE drivers). As shown in Figure 2e, consistent with prior survey findings \cite{zhang2025effect}, we find that EV drivers exhibit 26\% lower VMT than their ICE counterparts. 

Next, we examine charging session characteristics by comparing with EV Watts session-level data. We compare our EVCS subset (11,202 chargers) to EV Watts (1,116 chargers) and find that EV Watts' spatial coverage and type of chargers vary significantly by city (see Supplementary Figure S4). In the Bay Area and Seattle, the EV Watts samples are concentrated near high-traffic highway corridors. Meanwhile, in Boston and Denver, EV Watts chargers are more evenly distributed across the EVCS system, meaning more robust coverage of chargers on everyday city streets. Level 2 charger sessions compose a much smaller percentage of EV Watts sessions than our sessions in the Bay Area (0.0\% of EV Watts sessions, 58.5\% of our sessions) and Seattle (24.0\%, 71.8\%). Comparatively, in Boston (87.5\%, 78.2\%) and Denver (79.4\%, 67.9\%) Level 2 charger sessions are more highly represented in EV Watts sessions than our sessions. Given these differences, we compare the EV session characteristics separately by level two (L2) chargers and direct charging fast chargers (DCFCs). 

We then compare the average session duration of EV Watts charging events to our selected EV subset, and against the identified EV charging sessions for users that are not in our EV cohort. The EV Watts' duration is calculated as the total time the vehicle is plugged in, where our EVCS session duration is the time between the first and last observation within 10 meters of the EVCS. As a result our calculation is subject to more variance due to the heterogenous observation rate of device-level data (See Supplementary Figure S1). Even so, we find that the inferred EV driver cohort's median session length aligns with the EV Watts averages in each city for each charger type (Figure 2f-g), and the level-2 session duration is significantly longer than similar stops by non-EV drivers. Overall this indicates that the data processing filters out transient pings of ICE vehicles around EV charging stations. 

Comparing the weekday temporal charging profile of EV Watts' sessions to our identified EVCS sessions, and the overall stop profile of our users, we find distinct charging profiles for L2 EVCS and DCFC EVCS, further supporting our EV cohort. Our inferred EV driver cohort exhibits peak L2 public charging activity during weekday mid-mornings, with utilization tapering off throughout the afternoon (see Figure 2h). This pattern is distinct from general mobility stops, which show a heavy afternoon peak, and matches the EV Watts timing profile more closely. Survey findings on charging and driving behavior provide further verification. Survey findings indicate \cite{tal2020pev} that public charging peaks at 8:00 AM on weekdays, aligning with the temporal charging profile in the estimated EV driver subset. This behavior stands in sharp contrast to the evening-dominated peaks typical of residential charging \cite{tal2020pev}, where users plug-in upon returning home. 

Our cohort's demographic profile also matches established socioeconomic benchmarks for EV adoption \cite{lee2020exploring, javid2017pev, nazari2019ev}. As shown in Figure 2j, our EV cohort has a higher median income, lower poverty rates, a larger Asian population, and lower Black and Hispanic populations. However, there are two demographic trends that are not consistent across cities: home ownership and higher percent white population. Surveys show that EV owners have a higher rate of single-family home (SFH) ownership, but within EV owners SFH residents are typically the least likely to utilize public infrastructure \cite{funke2019much}. As our methodology captures drivers who use public stations, this under-representation of SFH owners is expected, and the lower proportion of percent white is reasonable as SFH ownership is skewed white in the United States due to a long history of racist housing practices~\cite{habibi2025quantifying}. Overall, the EV subset matches the main demographic trends and captures the mixed public charging behavior often observed in high-adoption urban areas.

\begin{figure}
    \centering
    \includegraphics[width=\linewidth]{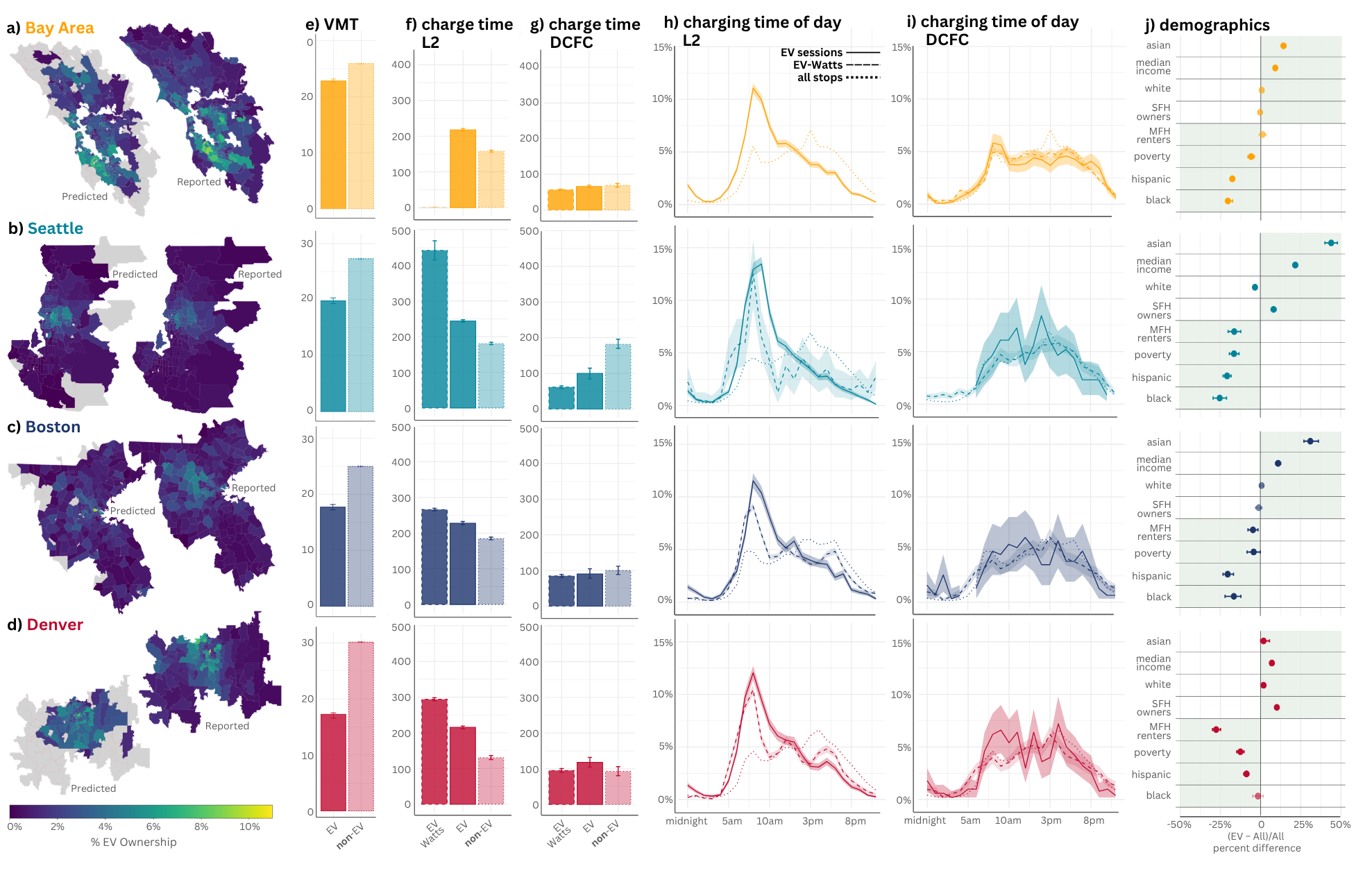}
    \caption{\textbf{The inferred EV cohort validates against expected traits across several behavioral and demographic dimensions.} \textbf{a-d)} Spatial distribution of inferred EV owners and state-reported registration numbers across four major MSAs . Correlation between zip level predictions and state-reported registrations range from 0.54-0.86 (Bay: 0.63, Seattle: 0.86, Boston: 0.54, Denver: 0.54).  \textbf{e)} Average daily Vehicle Miles Traveled (VMT) is lower for our identified cohort, consistent with the intensive-use patterns reported by surveys\cite{zhang2025effect}. \textbf{f)} Mean EV charging for locations with L2 chargers and no reported DCFC chargers. The Bay Area has no EV Watts data without DCFC plugs. Charging durations in our EV cohort are longer than charging durations for the non-EV cohort across all cities. \textbf{g)} Mean EV charging for locations with DCFC chargers and no reported L2 chargers. Charging durations in our EV cohort are shorter than or indistinguishable from charging durations for the non-EV cohort across all cities.  \textbf{h)} L2 EVCS charging session timings (solid) show a characteristic morning public charging peak that matches EV Watts session  timings (dashed) and is distinct from the late-afternoon peak of all mobility stops (dotted).  \textbf{i)} DCFC EVCS charging session timings (solid), EV Watts session timings (dashed), and all mobility stops (dotted), are distributed throughout the day, and do not show the early morning peak of L2 charging. \textbf{j)} Demographic signatures mirror established survey benchmarks, showing higher income, lower poverty, and characteristic shifts in housing type (SFH) and ethnicity for EV drivers compared to our entire study population.}
    \label{fig:placeholder}
\end{figure}

\subsection*{Evaluating EV driver behavior shows distinct activity trends} 

The inferred EV driver cohort enables a high-resolution analysis of activity patterns during public charging, as well as mobility outside of charging sessions. This creates opportunities for downstream applications, including quantifying economic spillovers across commercial categories, identifying mobility patterns of EV drivers to inform transit planning, and improving understanding of demand dynamics for grid management. 

Using the “stops” data, we examine behavior during charging sessions by classifying whether EV drivers stay near their vehicle, visit a nearby point-of-interest (POI), or leave their vehicle without visiting an identified POI. We find that these  patterns are consistent across cities, with 18-24\% of charging sessions involving a POI visit. To characterize these visits, we integrate SafeGraph POI data \cite{deweyPOI2022} and group destinations into eight categories: restaurants and cafes, beauty, fitness, pharmacy, grocery, entertainment, errands and retail. We define a POI visit as a stop within 50 meters of the POI and within 200 meters of the EVCS, resulting in 8,344 POI stops by an EV driver during an EV session out of the 4.5 million stops that occur near an EVCS.

We then compare the probability of visiting a POI by category for EVCS drivers to the broader study population. We find that EV drivers are significantly more likely to visit restaurants and gyms than the general public across the majority of the cities. Within these categories, cafes exhibit the largest relative increase in visitation. These results suggest that businesses located near EVCS, particularly cafes, may capture increased economic activity from charging-related foot traffic, aligning with findings that link EVCS presence to an increase in spending at restaurants \cite{zheng2024effects}.

\begin{figure}
    \centering
    \includegraphics[width=1\linewidth]{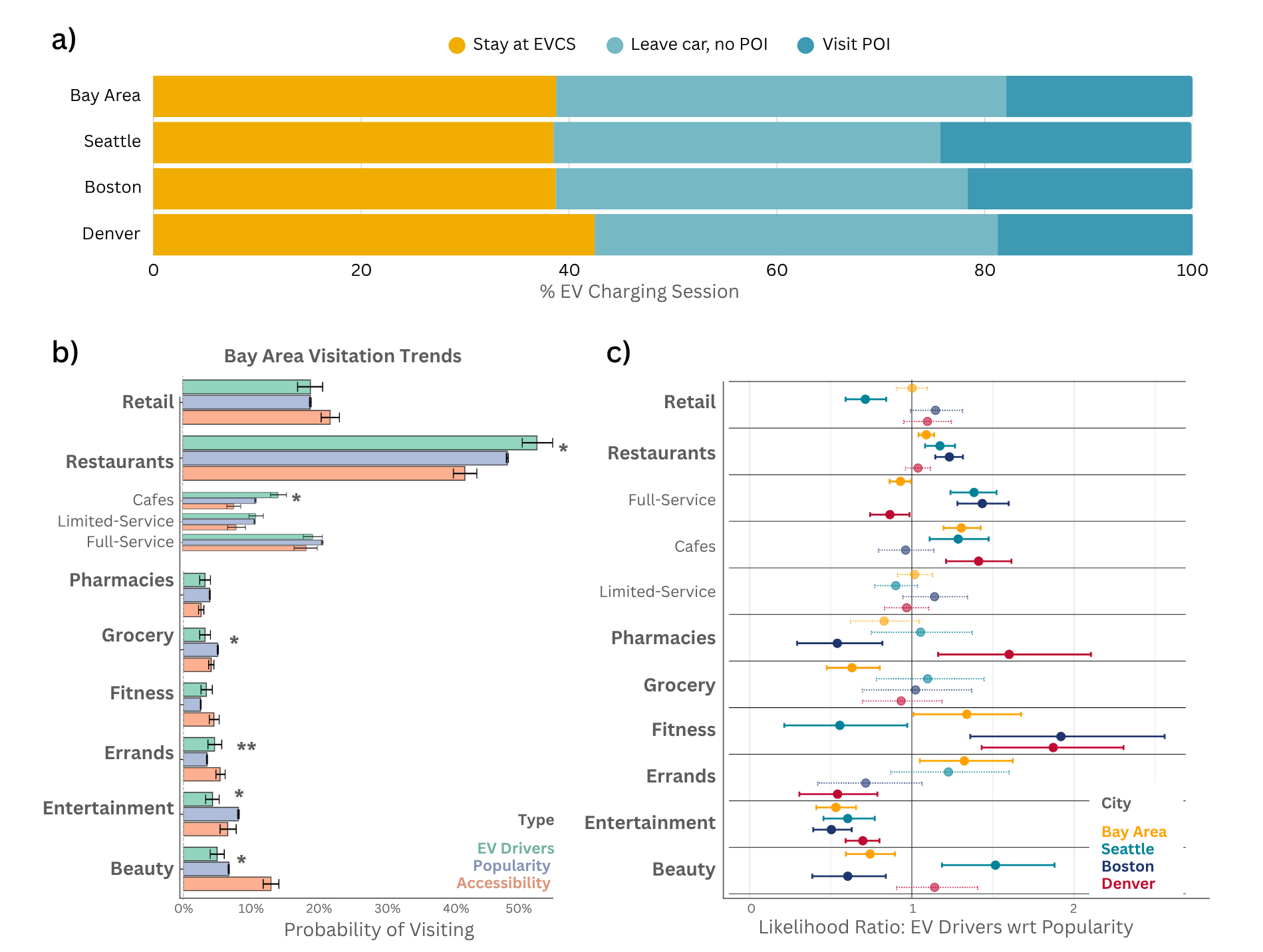}
    \caption{\textbf{EV driver POI visitation during charging is significantly different than overall POI visitation.} \textbf{a)} EV driver behavior during an EV charging session is calculated across all sessions of our identified cohort. Approximately 20\% of sessions are connected to a POI visit, across all cities. \textbf{b)} Probability of visitation by POI type in the Bay Area, given a POI visit within 250m of an EVCS.  Three groups labeled by color:  EV drivers (during a charging session), all users that visit a POI within the EVCS buffer zone (popularity), and the overall prevalence of each POI type (accessibility). EV drivers in the Bay Area show significantly higher visitation to restaurants, cafes and other errand category POIs during charging (see S8. Asterisk (\textbf{*}) indicates significant difference between EV driver and study population behavior ($p<0.01$), and (\textbf{**}) indicates ($p<0.05$). \textbf{c)} The visit likelihood ratio between the average EV driver and average popularity is calculated for each category where a value greater than one indicates a higher POI visitation likelihood for EV drivers. Color indicates the city, and solid lines indicate the probabilities are significantly different with a 95\% confidence interval. The Restaurant and Cafe categories display a consistent trend across cities.}

    \label{fig:fig3}
\end{figure}

Next, we extend the analysis beyond charging events to examine EV driver behavior over time. The inferred EV driver cohort allows us to evaluate whether EV drivers behave differently on charging days, non-charging days, and across the broader non-EV population. To ensure we are comparing users with similar observation rates, we restrict the analysis to days with sufficient observations. We define an active day as a day when a user has more than two observed stops, excluding home or work, and at least 36 pings. Within the EV driver cohort, we label days with an EVCS session as a ``charging day''. Using these definitions, we consider all POI stops that occur within the study area and compute the average daily POI visit rate. As shown in Figure 4a, we find that EV drivers visit more POIs on days when they are charging (2.0-2.2) compared to non-charging days (1.7-1.9), which suggest an association with the event of charging and users' overall travel behavior.  

To further characterize these behavior changes, we analyze the differences in POI visitation patterns including average stop duration, time between stops, and the radius of gyration of daily POI stops. In order to compare across groups that have different frequency of stops, we further subset the data to days where users have at least 2 POI stops. We then calculate the average stop duration and find that EV drivers have a significantly longer average stop duration during charging days indicating that EV drivers have both more frequent and longer stops on days with a charging event (Figure 4c). Next we evaluate how charging impacts the overall mobility pattern. We find that EV drivers have a shorter duration between POI stops (excluding the dwell time) and smaller radius of gyration of POI stops during days when they charge. These trends indicate trip bundling behavior on days that EV drivers charge, potentially showing that charging induces an ``errands day'' behavior where charging is grouped with other planned outings or errands.

\begin{figure}
    \centering
    \includegraphics[width=1\linewidth]{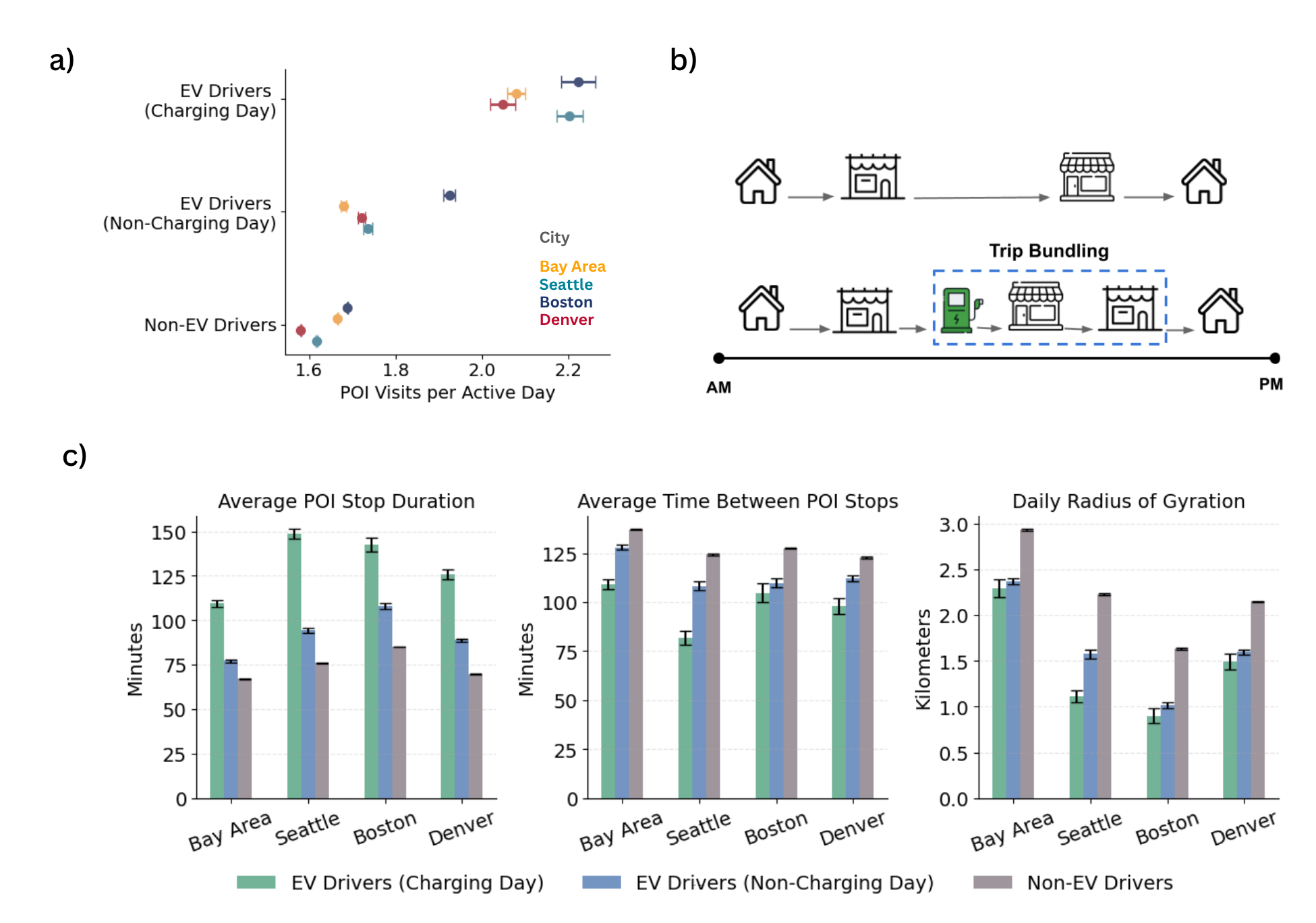}
    \caption{\textbf{EV drivers behave differently on days with charging events} \textbf{a}) The average daily POI visits (with 95\% CI) across three groups: EV drivers on days where they charge, EV drivers on days where they do not charge and all other users for days that meet the minimum activity requirements. \textbf{b}) Trip bundling is characterized by higher number of stops over less time and distance.  \textbf{c}) Left: Average stop duration by group for users who have at least two daily POI stops. Center: Average time between POI stops by group. Right: Radius of gyration for daily POI visits by group.
    }
    \label{fig:fig3}
\end{figure}

\section*{Discussion}\label{sec12}

Our method moves beyond the `black box' of aggregated registration data by using high-resolution mobility traces to identify EV drivers with high fidelity. This individual-level granularity allows us to analyze the specific behavioral drivers and planning patterns that aggregate data obscures. This allows us to provide data on their movements both on and off-plug. Additionally, the robust validation of our pooled model across individual EV charging data indicates transferability to other cities with only zip code level EV ownership data. As electric vehicles capture an increasing share of the global market and the EV landscape continues to evolve, understanding the intersection of charging infrastructure and human behavior become a prerequisite for resilient urban planning. 

We applied this inferred cohort to two primary case studies that underscore the utility of granular mobility data. First, our analysis of secondary behaviors during charging sessions reveals statistically significant visitation patterns during charging. We find that EV drivers are more likely to visit cafes and restaurants while tethered to a charger, demonstrating how the physical placement of chargers connects to economic spillover. These findings suggest that EVCS locations are not merely utility nodes but are drivers of foot traffic that can be leveraged to support local businesses. Second, we found significantly different behavior for EV driver daily POI visitation trends on charging days, indicating the EVCS charging environment has the ability to impact behavior throughout the entire day. On charging days, EV drivers exhibit trip bundling behavior while increasing both stay duration and frequency, amplifying total possible impact that optimizing EVCS environments may have on the local economy. These insights point to a significant opportunity to analyze how the physical environment shapes driver behavior during charging sessions. 

More broadly, our proposed EV identification method enables an entirely new data source at an unprecedented spatial and temporal scale. As the EV owner demographic is growing and changing, recent behavioral insights are vital. Especially as the EV subset is broadening from a majority white and wealthy cohort, data availability is imperative to capture evolving behavioral trends and provide equitable design and placement of EV charging infrastructure. There is a wide body of literature that establishes minority and non-SFH populations have less access to public EVCS infrastructure, while flagging the feedback loop that EV infrastructure availability drives EV adoption, and EVCS installment is based on EV adoption rates \cite{hsu2021public, canepa2019early}. By definition, our study population diverges from the full EV population as we can only detect users who use a public charging station. Providing a unique opportunity to understand who uses public chargers and facilitate equitable public charging infrastructure placement, potentially creating economic spillovers and increased social mixing at the same time. 

While our approach demonstrates the viability of identifying EV drivers through mobility traces, several limitations remain that could be addressed in future work. Most notably, our cohort selection relies on a top-down thresholding method: we calibrate our subset to match administrative registration totals, adjusted downward by a survey-derived estimate of drivers who never utilize public infrastructure. This dependence on static, region-specific surveys is a significant constraint, as there is currently a lack of granular data on public-versus-private charging practices outside of early-adoption hubs like California. This may further be exacerbated by plug-in hybrid vehicle (PHEV) owners, as we are not currently including PHEVs when setting expected ownership thresholds due to their differences in charging behavior. By using a fixed ownership threshold based on a country-wide survey, rather than a purely behavioral classification, we may be under-representing low-frequency public charger users. 

Our findings have implications far beyond the case studies we presented. Our proposed EV identification methodology increases the scale on which we can study research questions at a fine-grained, individual level across a wide variety of electric vehicle research and policy applications, including public EVCS infrastructure planning, grid management, and demand response.

\section*{Methods}\label{sec11}


\subsection*{Data}

We use an anonymized location dataset of mobile phones and smartphone devices provided by Cuebiq~\cite{cuebiq_data}, a location data intelligence company that collects anonymous, privacy-compliant location data of mobile devices using their software development kit (SDK) technology in mobile applications and privacy framework. Cuebiq processes data collected from mobile devices whose owners have actively opted in to share their location, and require all application partners to disclose their relationship with Cuebiq, directly or by category, in the privacy policy. With this commitment to privacy, the dataset contains location data for roughly 15 million daily active users in the United States. Individual level data analysis was done only within Cuebiq's Data Platform. All data analyzed in this study are aggregated to preserve privacy. In order to further preserve privacy, Cuebiq removes visits to sensitive locations in accordance with its Sensitive Points of Interest Policy. 

This location dataset allows us to implement a fine-grained analysis of individual movement patterns while maintaining privacy protections. To be included in our cohort, a user must have a home location within the counties of our study areas, have a median daily ping rate of at least 36 for at least one month in the study period (a ping every 30 minutes, excluding 8 hours to account inactivity due to sleep) and have at least 5 days of data (see Supplementary Figure S1 a and b). For users that meet those requirements, we include their data across the entire three month study period (see Supplementary Figure S1 c). Before we analyze user data,  we remove any pings that have a time difference below 3 seconds or a speed above 120km/h as the combination of proximity and accuracy at that scale created too much noise. Lastly, we only look at zip codes with at least 100 users that meet the above requirements (see Supplementary Figure S3). The resulting cohort includes 761,666 individuals (Bay: 232,662, Seattle: 179,155, Denver: 170,219, and Boston: 179,630) over a three month period from March 1st to May 31st, 2022. 

We selected the four cities in our study area using three criteria: high rates of EV adoption, availability of zip level EV ownership data and geographic diversity. We used the official Metropolitan Statistical Area (MSA) definitions for Denver, Seattle, and Boston. However, we do not adopt the MSA delineation for San Francisco, as it does not fully capture regional commuting patterns or all the counties with high EV ownership rates. Instead, we defined the study area to include all nine counties commonly recognized as part of the San Francisco Bay Area, allowing EV behavior dynamics to be seamlessly captured across multiple counties. Additionally, we elected to include the Bay Area with limited availability of EV Watts data for level 2 EVCS due to the high EV ownership rate and the availability of survey data and other relevant studies. 

We integrate data from several other sources: the U.S. Department of Energy's Alternative Fuels Data Center (AFDC)\cite{afdc_ev_locations} and Open Charge Map\cite{openchargemap_2025} provide us with locations of charging stations, and SafeGraph\cite{safegraph_global_places_2022} provide us with the locations and metadata on POIs. We construct our gas station location dataset using Safegraph POI data with the subcategory ``Gasoline Stations with Convenience Stores''.  We construct the EVCS location dataset using all data available in AFDC, Open Charge Map and Safegraph POIs under the subcategory ``Other Gasoline Stations'', taking care to not have duplicate records by spatially joining using a 5 meter buffer. To allow for the maximum chance of observing an EV charging session, we do not remove EVCS that are private or have membership restrictions, but instead remove EVCS that directly overlap with work or home locations. Cuebiq obscures user's home and work location by assigning a random point within the CBG and labeling it upleveled, we iterated through all the upleveled points in the dataset and removed those that were co-located within 5 meters of an EVCS. This prevents both identifying false EV charging sessions at home or work CBGs due to the randomly generated point being near an EVCS. In total, our dataset includes 11,202 EVCS (Bay: 6,761, Seattle: 1,557, Denver: 974, and Boston: 1,910), and 7,185 gas stations (Bay: 2,253, Seattle: 1,479, Denver: 1,352, and Boston: 2,101). Some sources identify points for each individual EVCS charging plug which we treat as a single station if they are within 100 meters (still analyzing visits at the port level), while gas stations are identified by the station in its entirety. 

We use several sources of data to validate our EV subset, including EV Watts data, zip code EV ownership data and ACS 5 year 2022 demographic data at the CBG level. Restricting our EV Watts data to the same three month period (March through May, 2022) across our four study areas, we have 1,116 total charging stations (Bay: 133, Seattle: 104, Boston: 554 and Denver: 325), 12,619 total users (Bay: 2,601, Seattle: 849, Boston: 5,744 and Denver: 3,425) and 76,480 total charging sessions (Bay: 13,164, Seattle: 5,463, Boston: 31,120 and Denver: 26,733). 

Zip code EV ownership data is not uniformly available across the U.S. and the availablity of that data served as a constraint. For our selected cities we found the data through publicly available government sources. Data for the Bay was sourced from the California Energy Commission, data for Seattle was sourced from state vehicle registration data, data for Boston was sourced from the MassDOT vehicle census, and data for Denver was sourced from the Atlas Public Policy dataset \cite{cadmv2019, wadol2026, massdot2023, atlas2024} . We also use the zip code level data to calculate our overall EV ownership rate for each study area.


\subsection*{Identifying EVCS Sessions}



To identify EVCS and gas station ``fueling sessions" we first limit our sample to ``slow points" (which we define as location pings with a movement speed below 10km/h).  Next, we identify slow points that are within 10 meters of an EVCS. Our sample includes 298,686 users with at least one EVCS slow point. The EVCS slow point can be the start or end of the session, or a point within the session.  We look at all points preceding the slow point until a point exceeds either a speed threshold of 20km/h or a distance threshold of 1km, and we do the same for the points after the slow point. Then we evaluate the continuous set of points, if there are two points within 10 meters (see Supplementary Figure S2 for 5 meter sensitivity analysis) of the EVCS with at least a duration of ten minutes (see Supplementary Figure S3 for 20 minute sensitivity analysis) we label that a charging session. The session duration is calculated as the time elapsed between the first EVCS ping within 10 meters and the last EVCS ping within 10 meters (see Figure \ref{fig:fig1}). We drop sessions with a duration less than 10 minutes. Using this definition, we identify 48,477 users with at least one EVCS session during the study period. 

To provide a behavioral counterpoint, we also identify gas station visits. Gas station visits are identified using the Cuebiq stop data, and SafeGraph POI data. Behavior during gas station visits is different. While the total duration is shorter, users will be at the station the entire time the car is fueling, which is much more likely to be captured by Cuebiq as a stop. We require a 15 meter proximity to the gas station centroid with a stay duration between 2 and 15 minutes to constitute a stop. We identify 324,592 users with at least one gas station session during the study period. 

\subsection*{Identifying EV Drivers}

The process of EVCS session identification presents several challenges. EVCSs are generally in highly trafficked locations, like mall parking lots, where presence near an EVCS does not necessarily indicate a charging event. For example, someone may be plugging in at the grocery store parking lot while they shop or they may simply be parking near an EVCS. As a result, our initial detection of EVCS sessions likely includes noise from non-EV drivers who are simply parked near chargers. 

The proportion of users with at least one detected EVCS session supports this concern. In the Bay Area, the EV ownership rate is approximately 4.4\%. Considering that 26.9\% of EV owners report never using public chargers \cite{turner2025ev}, we would expect to identify roughly 3.2\% of users as EV drivers.  While the Cuebiq sample is likely to over-represent EV drivers given their ``selection bias towards wealthier users", \cite{lake2024bias} we find 6.4\% of users have at least one EVCS session, which indicates some identified sessions are false positives. 

To distinguish true EVCS charging sessions from incidental proximity, we evaluate users' behavior longitudinally over the three month study period. For each user, we generate features capturing patterns related to EV ownership, including: the number of gas visits per month, the number of unique EVCS visited, if they have more EVCS sessions than gas station visits, time between EVCS sessions, how often they visit the same stations, the percentage of multi-family home housing in their home census tract, etc. Aside from the census variables, these are all generated solely from individual mobility data.

Many of the generated EV and gas related features are highly collinear. In order to select a parsimonious model, we use 4-fold cross validation to narrow down which features were strongly predictive of the out of sample values. Our top performing models had comparable performance, and as a result we selected our final model based on the interpretability of the coefficients. 

Our final model includes: city, gas stops per month, VMT, percentage of multi-family housing, unique EVCS visited, and a binary indicating if EVCS visits exceed gas station visits. We train a logit regression model on data pooled across all study areas to predict the zip code level EV ownership rate, as reported by local government agencies. While this outcome variable does not represent individual level EV ownership, it provides the most granular publicly available measure of EV ownership rates, allowing the model to learn feature patterns associated with areas that have higher rates of EV adoption. 


To select our EV driver cohort, we apply the model's predictions at the individual level. We find the percentage of EV ownership for the Bay Area(4.4\%), Seattle(1.9\%), Boston(1.8\%) and Denver(5.5\%), adjust it by the 26.9\% that never use L2 public charging and then classify that top percent of users with the highest predicted probabilities as EV drivers.  Our final EV cohort consists of 19,228 EV drivers across the four cities with 7,512 in the Bay area, 2,450 in Seattle, 2,405 in Boston  and 6,861 in Denver. This approach calibrates the share of EV drivers in our cohort with the known ownership rates for each metropolitan area, while selecting the users most likely to represent true EV owners.

\subsection*{Baseline Models for EV Driver Estimation}

We use two baselines. First, our heuristic baseline classifies anyone with one or more EVCS sessions and no gas station visits over the three month period as an EV driver. We selected these parameters to represent a scenario where any EVCS session indicates an EV driver as long as they have no gas station visits. This heuristic model does not consider false positive EVCS sessions, but if we increase the threshold to at least two charging session, our overall EV ownership rate is below 1\%. This method doesn't calibrate to the expected number of EV drivers in each metro the way our model based methods do. This method gives an overall EV ownership rate of 2.7\% (Bay Area: 3.5\%, Seattle: 2.9\%, and Boston: 1.7\%, Denver: 2.5\%) and has a correlation of 0.40. We further use survey data and EV Watt to validate the EV subset (see Supplementary Figure S5), finding VMT, duration of charge and demographics to not align as expected. The temporal trends align more closely, but this may partially be due to the fact that we are sampling from very similar subsets since only 6.4\% of all users have an EVCS stop. 

Second, our model baseline functions similarly to our main model, predicting zip level EV ownership and selecting the EV cohort using the top percent that corresponds to the city's overall EV ownership. The model only includes city, individual VMT, and the median income of each individual's home Census tract. This model has a correlation of 0.55. Although the correlation is decent, the spatial mismatch shows overestimation in certain tracts. Since we use median income and VMT to select the data, we see good alignment with the expected demographics and VMT trend. However, for all other validation at the individual level, we do not see the expected alignment (see Supplementary Figure S6).

\subsection*{Validation using EV Watts Data}

To evaluate whether our inferred EV cohort accurately represents real world behavior, we used the EV Watts database, a nationwide repository of session level charging data. EV Watts offers a significantly richer view of charging sessions than mobility traces, including data on pricing, State of Charge (SoC), and the distinction between active charging time and total dwell time. Additionally, the population it covers is broader than our mobility dataset, given their information on private charging as well. To adjust for this discrepancy, we processed the EV Watts data to ensure the behaviors we were measuring were as similar as possible across both sources. We trim the high-resolution metrics of EV Watts to match the constraints of our mobility-derived cohort. 

To account for our method's inability to capture private residential or depot charging, we filtered the EV Watts stations to exclude ports at single-family homes, multi-family homes, those reserved for fleets, and those labeled ``limited'' and ``private''. This subsetting ensures we are comparing the same types of chargers in both datasets, given there are different behavior patterns typical of public infrastructure.

An additional challenge in using session level data is the fragmentation of charging sessions. A single charging event often appears as several separate sessions if a charger technical issue occurs or a user has to re-authorize mid-charge. We combined consecutive sessions by the same account if they happened at the same location within 15 minutes of each other. As a result we did not include 7,098 (9.3 \%) sessions not linked to a unique user ID (Bay: 966 (7.3 \%), Seattle: 1,343 (24.6 \%), Boston: 1,636 (5.3 \%) and Denver: 3,153 (11.8 \%)), in the EV Watts sample. This grouping is essential for estimating total visit duration; since mobility data only shows us when a user arrives and leaves the vicinity of a charger, combining these fragmented events gives us a comparable sense of their dwell time. Without this adjustment, short interruptions would artificially deflate session lengths and inflate visitation counts.


Finally, we applied a duration filter to match our main identification method, keeping only sessions that lasted between 10 minutes and 24 hours. After processing, we have 56,872 total charging sessions (Bay: 11,150, Seattle: 4,001, Boston: 24,841 and Denver: 16,880) in the EV Watts data, 74\% of the original sample (Bay: 85\%, Seattle: 73\%, Boston: 80\%, and Denver: 63\%).

After we have our final EV Watts EV session subset, we compare the volume and spatial coverage to our EVCS session data (see Supplementary Figure S4). We find discrepancies between our data and EV Watts' data with regards to the distribution of EV charging sessions by charger type. We found that our dataset is much more skewed to level two charging with only 5-10\% (Bay: 10.9\%, Seattle: 6.0\%, Boston: 6.2\%, Denver: 5.2\%) of all sessions occurring at DCFC, compared to a much wider range in the EV Watts data (Bay: 100\%, Seattle: 76.0\%, Boston: 12.5\%, Denver: 20.6\%). Part of this discrepancy in DCFC charging session volume could be that we only look at charging session by drivers who reside in the study area and DCFC chargers often are used on longer trips \cite{sun2016fast}. In the EV Watts data the charger type is linked to each session so there is no ambiguity in session type.  However, our subset of chargers are sometimes defined at the plug level and sometimes defined at the station level, so we are not always able to identify if the charging session is linked to a DCFC or L2 plug. As a result, we classify all sessions at a station with at least one DCFC as such, and all sessions at a station with at least one L2 as such. This means that sessions can be classified under both categories. Of the DCFC sessions, we find that 26.0\% of them are at stations with both charger types (Bay: 24.1\%, Seattle: 26.9\%, Boston: 43.8\%, Denver: 8.7\%). Out of the L2 sessions, we find that 3.1\% of them are at stations with both charger types (Bay: 4.5\%, Seattle: 2.3\%, Boston: 3.5\%, Denver: 0.7\%).  Our subset of chargers sometimes do not have a charger type designation at all. In our charge time and time of day analysis, 33,600 sessions (26.3\%) are dropped for having no speed designation, and 161 (0.1\%) sessions are dropped for only having L1 chargers.

\subsection*{Analysis of POI Visits During EV Charging Sessions}
To understand the behavior of EV drivers during their charging session we consider their Cuebiq detected stops during the duration of their charging event. We consider all the stops within 200 meters of the charging station. We see four types of behavior: 1) EV drivers that stay in or near their car, 2) EV drivers that visits a POI, 3) EV drivers that leave their car but do not visit a POI, and 4) users that are not in the stop data. The first three groups were expected, but the unobserved group indicates that there are EV charging sessions  occurring that we can only see when using the device-level data (SF 46.8\%, Seattle: 50.1\%, Boston 50.0\%, Denver: 53.6\%). 

To understand this behavior, we consider the device-level data points during the charging session and calculate the distance from the EVCS for each observed point. We look at the 75th percentile of distance to make sure observations near the EVCS do not dominate other behavior while avoiding data outliers that can occur at the ping level. We find that 42.9\% (SF 43.5\%, Seattle: 42.8\%, Boston 41.2\%, Denver: 42.8\%) are within 10m and the other observations have an average of 80.3m ( SF 83.5m, Seattle: 71.4m, Boston 72.3m, Denver: 87.6m). We assign each session that had no ``stops" to it's respective category using ping data, and end up with three distinct charging session behaviors.

For an EV driver to have a POI visit, the stop must be within 50 meters of a POI in our dataset.  We create our POI dataset from a subset of SafeGraph data, creating  8 POI categories: cafes, beauty, fitness, pharmacy, grocery, entertainment from SafeGraph assigned top and sub categories. 

\begin{itemize}
    \item \textbf{Fitness \& Wellness:} Fitness and Recreational Sports Centers
    \item \textbf{Healthcare \& Pharmacies:} Pharmacies and Drug Stores; Health and Personal Care Stores
    \item \textbf{Beauty:} Personal Care Services; Hair Salons
    \item \textbf{Restaurants \& Caf\'es:} Restaurants and Other Eating Places; Drinking Places (Alcoholic Beverages); Food Services and Drinking Places; Special Food Services; Bakeries and Tortilla Manufacturing
    \item \textbf{Retail:} Clothing Stores; Shoe Stores; Department Stores; Clothing and Clothing Accessories Stores; Sporting Goods, Hobby, and Musical Instrument Stores; Jewelry, Luggage, and Leather Goods Stores; Book Stores and News Dealers; Electronics and Appliance Stores; Furniture Stores; Home Furnishings Stores; Used Merchandise Stores; Office Supplies, Stationery, and Gift Stores; General Merchandise Stores (including Warehouse Clubs and Supercenters); Other Miscellaneous Store Retailers; Florists
    \item \textbf{Grocery:} Grocery Stores; Grocery and Food Stores; Supermarkets; Specialty Food Stores; Food and Beverage Stores; Beer, Wine, and Liquor Stores
    \item \textbf{Other Errands:} Drycleaning and Laundry Services; Automotive Repair and Maintenance; Consumer Goods Rental; Personal and Household Goods Repair and Maintenance; Postal Service
    \item \textbf{Entertainment:} Museums, Historical Sites, and Similar Institutions; Amusement Parks and Arcades; Other Amusement and Recreation Industries; Motion Picture and Video Industries; Performing Arts Companies; Arts, Entertainment, and Recreation
\end{itemize}

Any POIs that are related to medical businesses or education are not included for privacy reasons. For the majority of POIs we use the centroid point, since the provided polygons often overlap or include the larger footprint of the building the POI is in, but for parks and museums we use the polygon to better capture the larger area they cover. To be considered a POI stop, the observed point must be within 50 meters of  the centroid or intersect with the poi polygon. 

After we classify POI visits, we want to understand visitation preferences of EV drivers by POI type. For each driver, we consider all EV charging sessions where a POI is visited, and calculate the proportion of visits by each type, creating a probability of visiting each type of POI. Next we want to understand if the POI visits by category are the same as the general population, or alternatively a product of the available POIs near EVCS. To create a comparable non-EV driver subset we look at all users with at least one POI visit within 250m of the EV charging stations visited by the EV driver subset. We then calculate the POI probability of visiting each type of POI for all users. As a result, we compare visits across the same subset of POIs and use Welch's t-tests to identify categories with significant differences between EVC and non-EVC groups. We also calculate spatial availability by counting the number of POIs in each category that are within 250m of the EVCS.

We then expand our area of evaluation and look at all POI visits across all users, considering three groups: non-ev drivers, ev drivers on a day where they have a identified charging session and ev drivers on days when we do not observe a charging session. We first limit our subset to days where users have at least 36 observed pings and more than two observed stops (excluding home, work and the EV stop). Then we calculate the total number of visits to POIs in our 8 categories normalized by the number of days they are active.  To evaluate stop trends throughout the course of the day, we further subset the data to those who make at least two POI stops so we can evaluate how the proximity in both time and space between stops is impacted by the presence of a charging event. Next we calculate the average daily radius of gyration for the POI stops weighted by the dwell time at each POI, finding that the effective distance traveled between visited POIs is smaller for EV drivers on charging days.

\bibliography{sn-bibliography}

\section*{Acknowledgements}

T.Y. and J.Y.J.C. acknowledge support by the National Science Foundation under grant number BCS 2425021. 

\section*{Author contributions statement}

C.C., A.D., M.C., J.Y.J.C., and T.Y. designed the algorithms, performed the analysis, and developed models. M.C., J.Y.J.C., and T.Y. supervised the research. All authors wrote the paper. The company data were processed by C.C. and T.Y. All authors had access to aggregated (non-individual) processed data. All authors reviewed the manuscript. 

\section*{Data availability}
The data that support the findings of this study are available from Cuebiq through their Social Impact program, but restrictions apply to the availability of these data, which were used under the license for the current study and are therefore not publicly available. Information about how to request access to the data and its conditions and limitations can be found in \url{https://cuebiq.com/social-impact/}.  
Data access requests should be submitted through Cuebiq's Social Impact customer page \url{https://cuebiq.com/demo/}, where the Sales team at Cuebiq may be contacted in a timely manner. 
EV Watts data were obtained from Livewire Data Platform funded by U.S. Department of Energy, Office of Energy Efficiency and Renewable Energy's Vehicle Technologies Office operated and maintained by Pacific Northwest National Laboratory, Idaho National Laboratory and National Renewable Energy at https://livewire.energy.gov.
Data about the points-of-interest locations was provided by Safegraph, who can be contacted through \url{https://www.safegraph.com/}. The Safegraph data is available through the Dewey platform through a paid subscription \url{https://app.deweydata.io/home}. 
Tiger shapefiles can be downloaded from the US Census Bureau \url{https://www.census.gov/programs-surveys/geography/guidance/tiger-data-products-guide.html}.

\section*{Code availability}
The analysis was conducted using Python. Code to reproduce the main results in the figures from the aggregated data is publicly available on GitHub \url{https://github.com/callieclark/EV_driver_patterns}.

\section*{Competing interests statement}

The authors declare no competing interests.

\renewcommand{\thefigure}{S\arabic{figure}}
\setcounter{figure}{0}
\renewcommand{\thefigure}{S\arabic{figure}}
\renewcommand{\thetable}{S\arabic{table}}

\end{document}


\date{} 

\maketitle





\setcounter{figure}{0}
\setcounter{table}{0}







\begin{table}
\centering
\begin{tabular}{lccccc}
\toprule
& \multicolumn{5}{c}{\textbf{Individual Cities}} \\
\cmidrule(lr){3-6}

& \textbf{Pooled Final Model} & Bay & Seattle & Boston & Denver \\
\midrule

Intercept
& $-4.2996^{***}$ & $-3.3722^{***}$ & $-4.3172^{***}$ & $-4.282^{***}$ & $-3.2129^{***}$ \\
& (0.000) & (0.000) & (0.000) & (0.000) & (0.000) \\

City (Denver)
& $1.1142^{***}$ & $-$ & $-$ & $-$ & $-$ \\
& (0.000) &  &  &  &  \\

City (SF)
& $0.9159^{***}$ & $-$ & $-$ & $-$ & $-$ \\
& (0.000) &  &  &  &  \\

City (Seattle)
& $-0.0459$ & $-$ & $-$ & $-$ & $-$ \\
& (0.134) &  &  &  & \\

EV Sessions $>$ Gas Stops
& $0.1138^{*}$ & $0.0762$ & $0.2813^{*}$ & $0.1438$ & $0.0655$ \\
& (0.0258) & (0.281) & (0.023) & (0.359) & (0.591) \\

Gas Stops/Month
& $-0.0392^{***}$ & $-0.0479^{***}$ & $-0.0451^{*}$ & $-0.0326$ & $-0.0325^{***}$ \\
& (0.000) & (0.000) & (0.020) & (0.079) & (0.000) \\

VMT (in 100's)
& $-0.1479^{***}$ & $-0.1033$ & $-0.4735^{***}$ & $-0.3537^{**}$ & $-0.0332$ \\
& (0.000) & (0.126) & (0.000) & (0.008) & (0.603) \\

\% Multifamily
& $-0.0464$ & $-0.0964^{*}$ & $0.1525^{*}$ & $-0.0174$ & $-0.0912^{*}$ \\
& (0.068) & (0.020) & (0.032) & (0.793) & (0.037) \\

Unique EVCS Stations
& $0.0634^{**}$ & $0.0526$ & $0.0914^{*}$ & $0.0773$ & $0.0572$ \\
& (0.004) & (0.065) & (0.037) & (0.348) & (0.422) \\

\midrule
R-squared
& 0.323 & 0.020 & 0.041 & 0.016 & 0.015 \\

Adj.\ R-squared
& 0.323 & 0.020 & 0.041 & 0.016 & 0.015 \\

BIC
& 44,382.32 & 18,187.36 & 4,904.42 & 5,124.59 & 16,342.19 \\

N
& 761,666 & 232,662 & 179,155 & 179,630 & 170,219 \\

\bottomrule
\end{tabular}
\caption{Regression coefficients for our pooled model, and for each city individually.}
\label{table:modeltable}
\end{table}

\begin{table}
\centering
\begin{tabular}{lccccc}
\toprule
& \multicolumn{5}{c}{\textbf{Individual Cities}} \\
\cmidrule(lr){3-6}

& \textbf{Pooled Baseline Model} & Bay & Seattle & Boston & Denver \\
\midrule

Intercept
& $-5.1530^{***}$ & $-4.4215^{***}$ & $-5.3751^{***}$ & $-5.1575^{***}$ & $-3.8386^{***}$ \\
& (0.000) & (0.000) & (0.000) & (0.000) & (0.000) \\

City (Denver)
& $1.1725^{***}$ & $-$ & $-$ & $-$ & $-$ \\
& (0.000) &  &  &  &  \\

City (SF)
& $0.7843^{***}$ & $-$ & $-$ & $-$ & $-$ \\
& (0.000) &  &  &  &  \\

City (Seattle)
& $-0.0082$ & $-$ & $-$ & $-$ & $-$ \\
& (0.791) &  &  &  & \\

VMT (in 100's)
& $-0.2701^{***}$ & $-0.2112^{**}$ & $-0.5430^{***}$ & $-0.6093^{***}$ & $-0.1471^{*}$ \\
& (0.000) & (0.001) & (0.000) & (0.000) & (0.018) \\

Median Income (in 10,000's)
& $0.0645^{***}$ & $0.0668^{***}$ & $0.0847^{***}$ & $0.0706^{***}$ & $0.0508^{***}$ \\
& (0.000) & (0.000) & (0.000) & (0.000) & (0.000) \\

\midrule
R-squared
& 0.532 & 0.402 & 0.398 & 0.291 & 0.234 \\

Adj.\ R-squared
& 0.532 & 0.402 & 0.398 & 0.291 & 0.234 \\

BIC
& 43,478.150 & 17,709.695 & 4,868.778 & 4,853.424 & 16,121.429 \\

N
& 735,391 & 223,570 & 175,674 & 169,468 & 166,679 \\

\bottomrule
\end{tabular}
\caption{Regression coefficients for our pooled baseline model, and for each city individually.}
\label{table:baselinetable}
\end{table}


\begin{table}[h]
\centering
\begin{tabular}{ || c c c c c || }
 \hline
 Sample & Bay & Seattle & Denver & Boston \\ [0.5ex] 
  \hline\hline
Everyone observed & 783,137 & 491,753 & 491,753 & 713,993 \\
$\geq$ 36 median points & 614,855 & 373,099 & 373,099 & 537,356 \\
$\geq$ 36, has home location in study area & 234,959 & 181,619 & 173,124 & 182,384  \\
$\geq$ 36, has home, $\geq$ 100 in zip, census data & 232,662 & 179,155 & 170,219 & 179,630 \\
 \hline
\end{tabular}
\caption{Our population sizes}
\label{table:3}
\end{table}

\begin{figure}
    \centering
    \includegraphics[width=\linewidth]{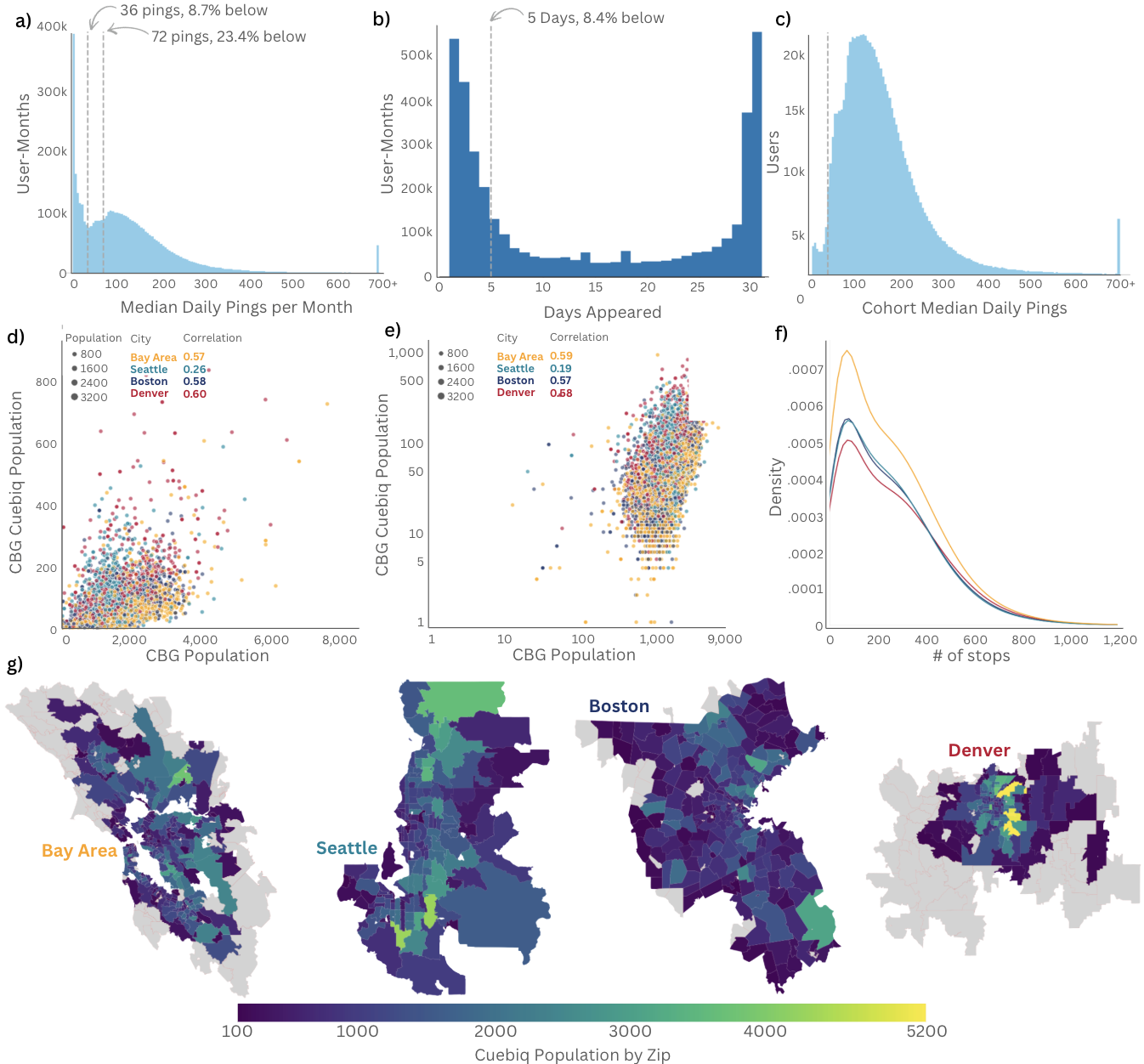}
    \caption{\textbf{Cuebiq data: Descriptive} \textbf{a)} The distribution of median number of daily pings over a month. Users are represented for each month they are in the data. We find ~8.7\% of user-months below our 36 ping minimum, and ~23.4\% of our user-months below our 72 ping minimum.  \textbf{c)} The number of days users appear in the data over each user-month. Users are most likely to show up on only a few days or for the entire month. We require that in one month  a user is observed for at least 5 days  and has a median daily ping rate of at least 36 to be included in our user subset. \textbf{c)} Median daily ping distribution of selected users across all included months. There is a small sample of users with overall median daily pings below 36 due to including all data on a user when the one month threshold is met. \textbf{d)} The relationship between Census block group population and the number of our Cuebiq users. \textbf{e)} The log-log relationship between Census block group population and the number of our Cuebiq users. \textbf{f)} The density of user-stops across cities. \textbf{g)} The spatial distribution of our final Cuebiq panel. }
    \label{fig:user_subset}
\end{figure}


\begin{figure}
    \centering
    \includegraphics[width=\linewidth]{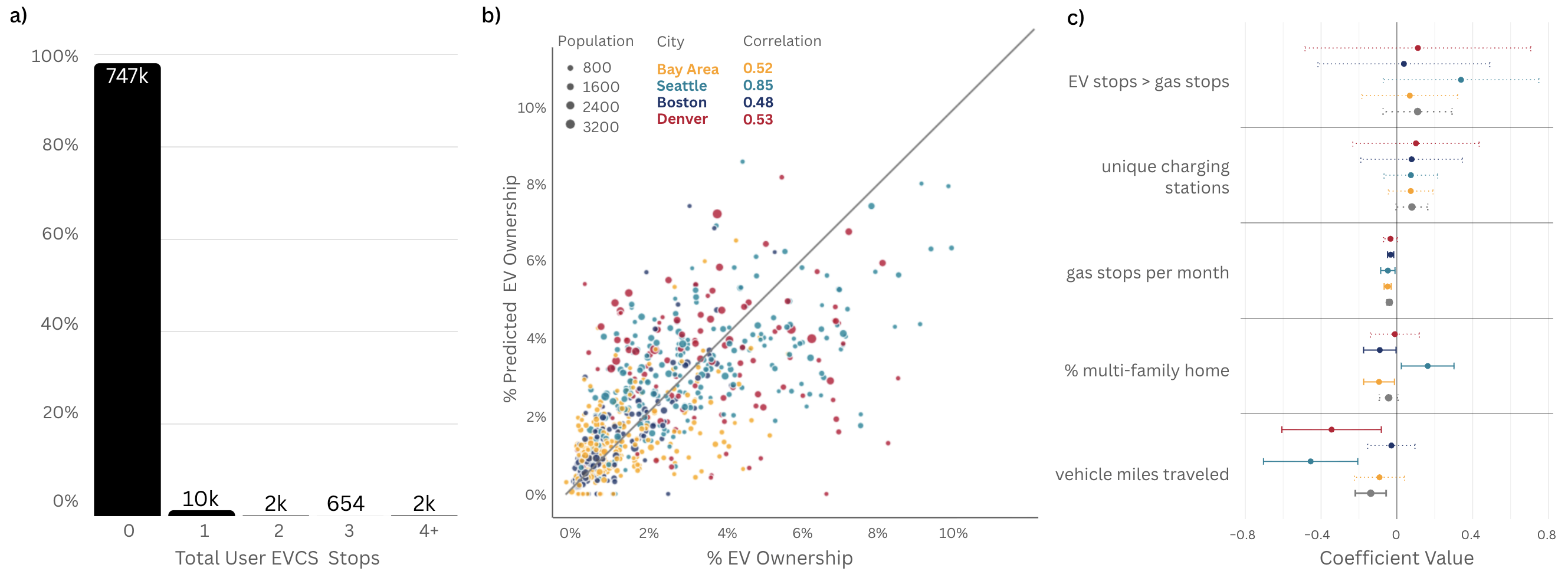}
    \caption{\textbf{Sensitivity testing: lowering the maximum distance from an EVCS from 10 meters to 5 meters.} \textbf{a)} EV sessions are less common, with 98.1\% of the cohort having no EV sessions, up from 93.6\%. \textbf{b)} The pooled EV cohort has an overall correlation of 0.59, down from 0.73. \textbf{c)} Model coefficients maintain directionality, but no EV related variables are significant.}
    \label{fig:fig6}
\end{figure}

\begin{figure}
    \centering
    \includegraphics[width=\linewidth]{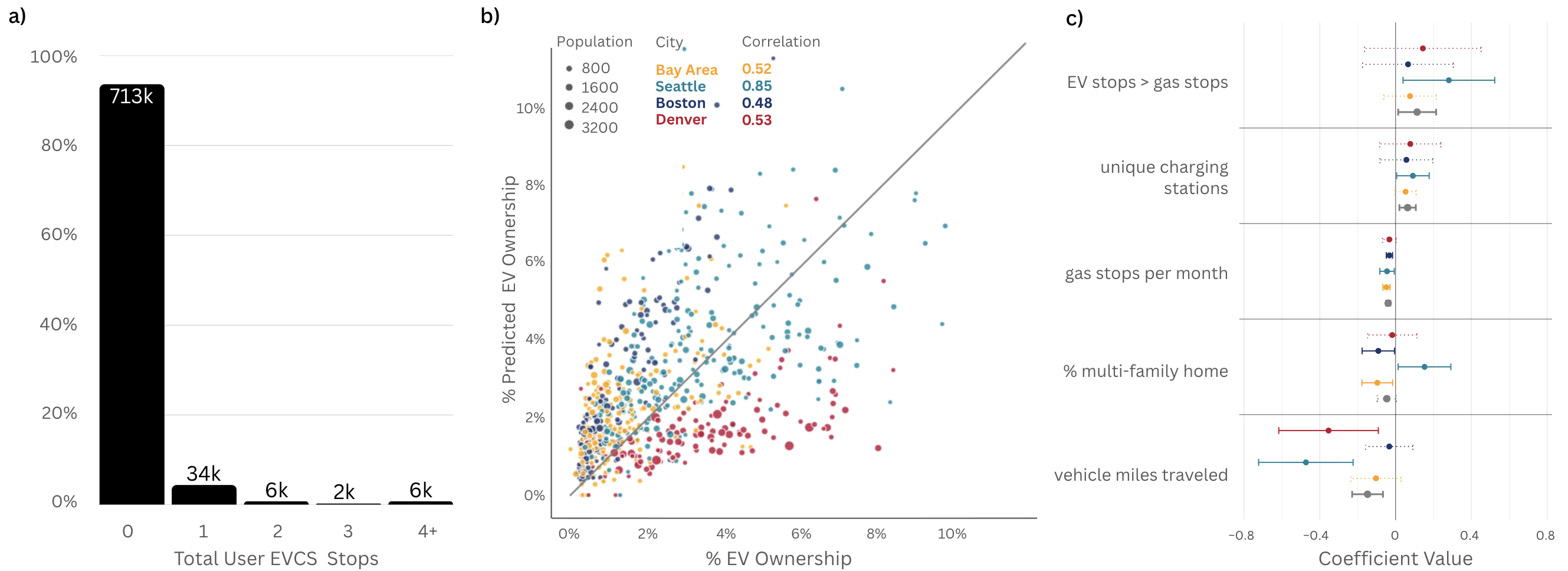}
    \caption{\textbf{Sensitivity testing: raising the minimum EVCS session length from 10 minutes to 20 minutes.} \textbf{a)} EV sessions are functionally equivalent to our main methods. \textbf{b)} The EV cohort has an overall correlation of 0.68, down from 0.73. \textbf{c)} Model coefficients maintain directionality, but EV related variables are largely insignificant. Seattle is the only city with significant EV related coefficients. }
    \label{fig:fig7}
\end{figure}


\begin{figure}
    \centering
    \includegraphics[width=\linewidth]{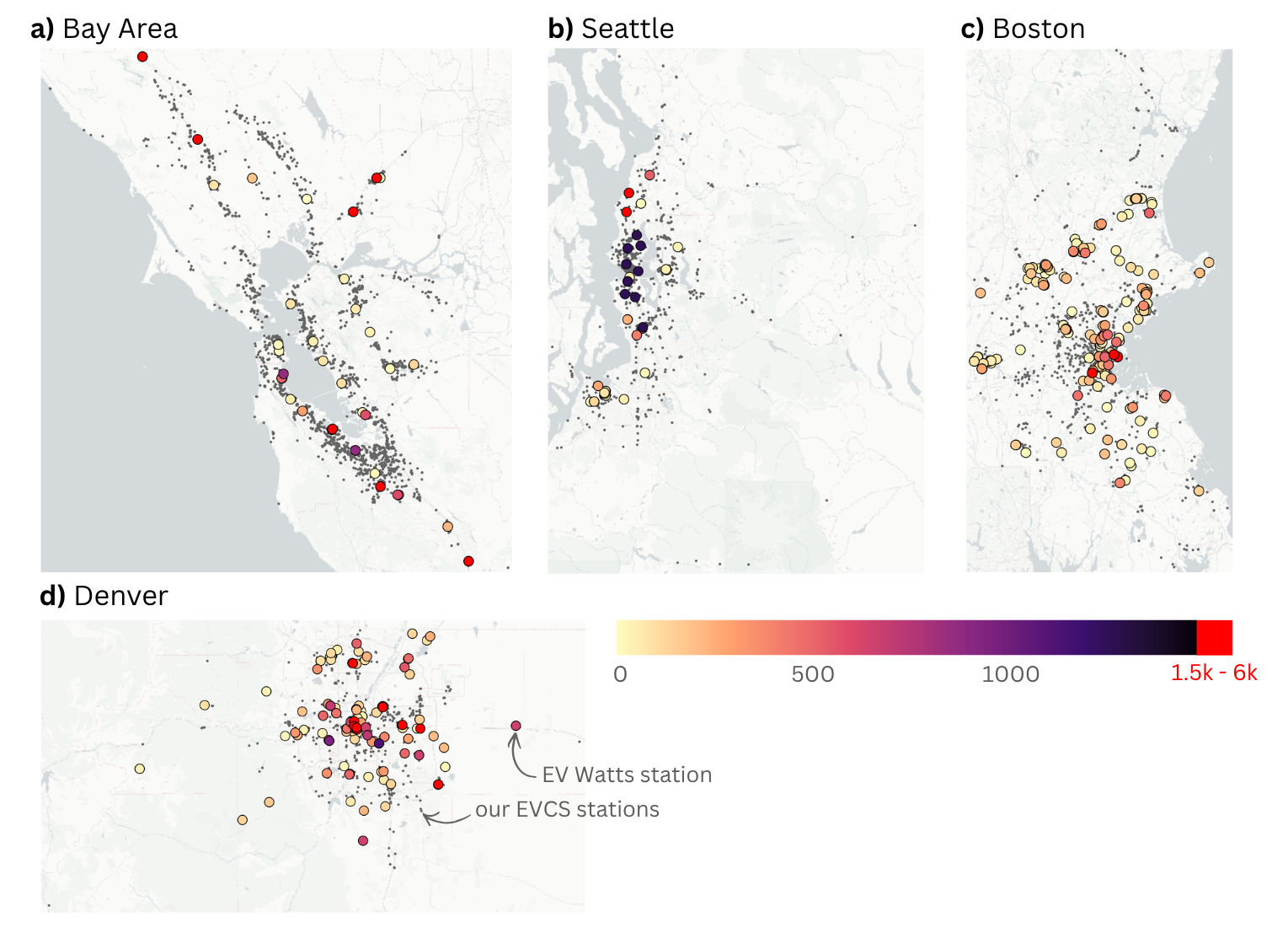}
    \caption{\textbf{EV Watts EVCS distribution and use comparison.} \textbf{a-d)} Compares our study EVCS spatial locations (light gray, no usage indicated) to EV Watts station locations (color determined by number of EVCS sessions, see legend).  }
    \label{fig:s_evwatts}
\end{figure}


\begin{figure}
    \centering
    \includegraphics[width=\linewidth]{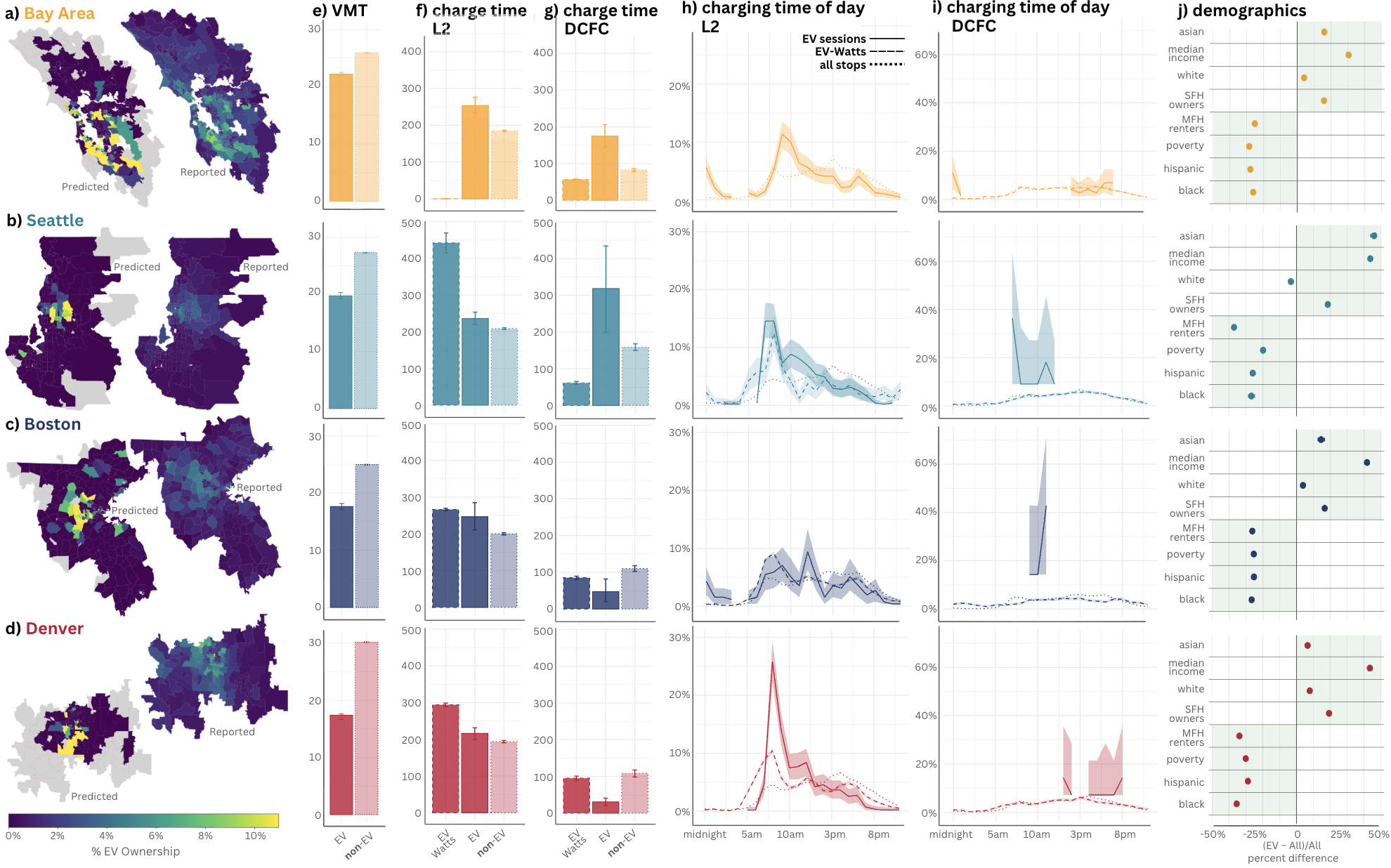}
    \caption{\textbf{Cohort validation for baseline model.} \textbf{a-d)} Spatial distribution of inferred EV owners is poorly correlated with state-reported registration numbers across four major MSAs. \textbf{e)} Vehicle Miles Traveled (VMT) have significant differences, which is expected given that VMT is a major predictor in the baseline model. \textbf{f-g)} Charging durations don't have good alignment with EV Watts ground-truth data. \textbf{h-i)} Temporal activity shows less consistent patterns across cities, and notably sparse data for DCFC charging sessions. \textbf{j)} Demographic trends match surveys, with median income being significantly higher than the general population as compared to our EV subset. Median income is highly correlated with other demographics so the overall alignment is expected.}
    \label{fig:fig8}
\end{figure}


\begin{figure}
    \centering
    \includegraphics[width=\linewidth]{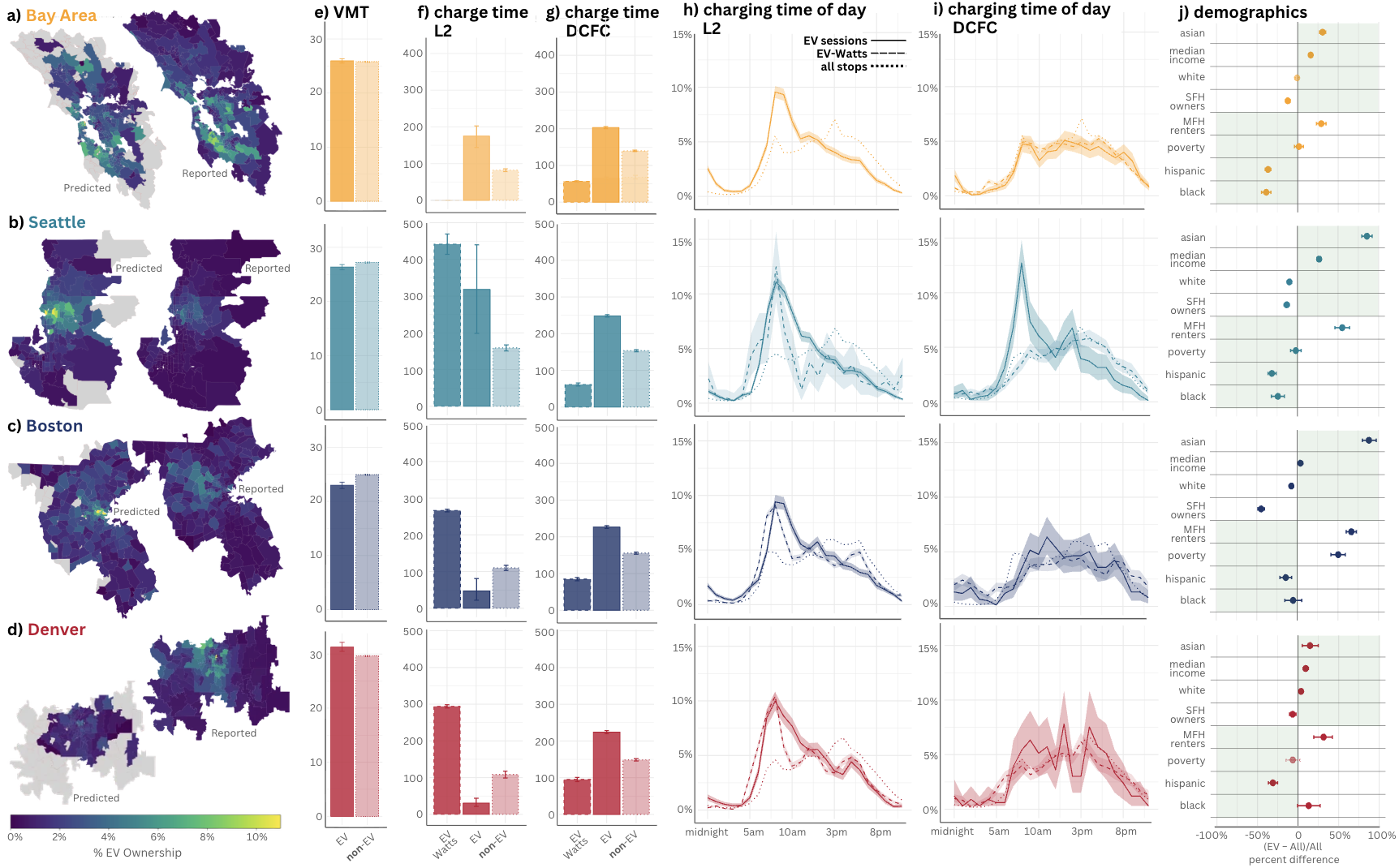}
    \caption{\textbf{Cohort validation for baseline heuristic.} \textbf{a-d)} Spatial distribution of inferred EV owners is correlated with state-reported registration numbers across four major MSAs. Compared to our EV subset, we see over-prediction in areas with dense EVCS infrastructure. \textbf{e)} Vehicle Miles Traveled (VMT) are fairly similar between the EV subset and all drivers, with small differences in both directions (Seattle and Boston are smaller). \textbf{f-g)} Charging durations don't have good alignment with EV Watts ground-truth data. DCFC charger durations are notably longer than both the EV Watts subset and the EVCS sessions of those who were not classified as EV drivers.  \textbf{h-i)} Temporal activity shows an even stronger morning peak.  \textbf{h)} Demographic trends don't match surveys across the board.}
    \label{fig:fig9}
\end{figure}

\begin{figure}
    \centering
    \includegraphics[width=\linewidth]{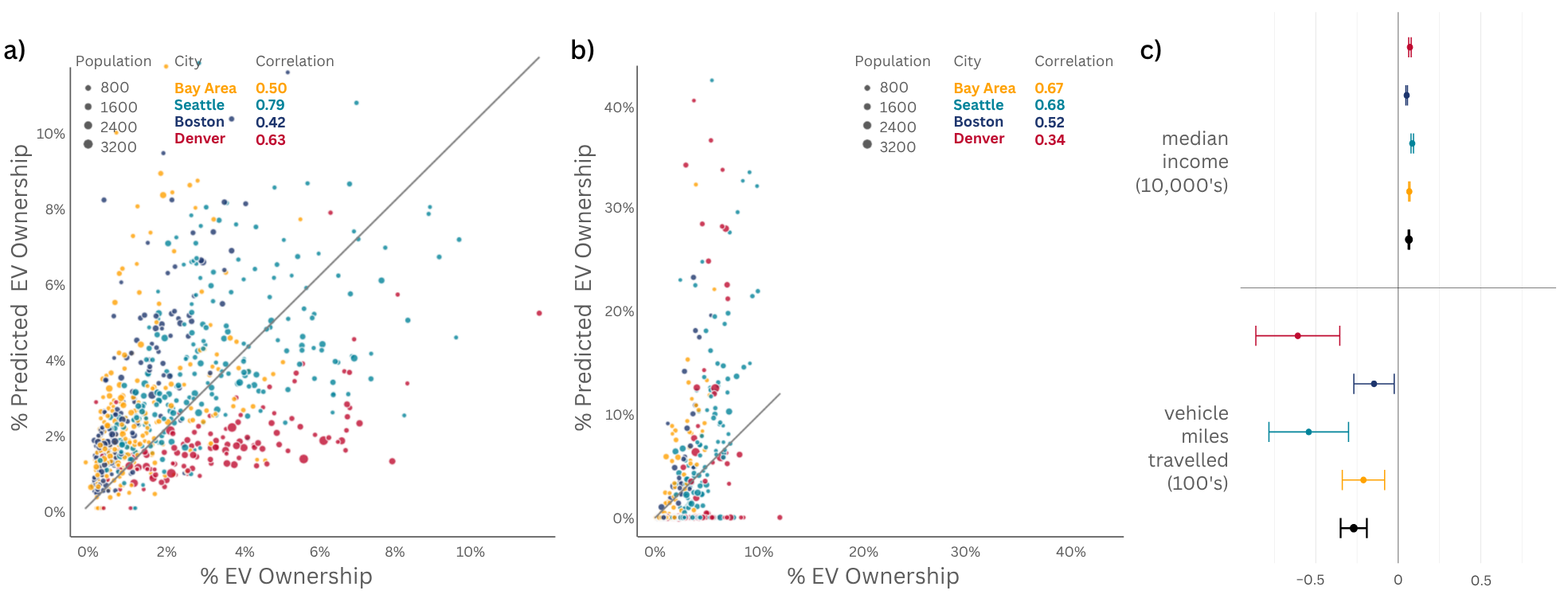}
    \caption{\textbf{Performance of baselines at the zip level.} \textbf{a)} Zip level EV prevalence of baseline heuristic as compared to ground truth. \textbf{b)} Zip level EV prevalence of baseline model as compared to ground truth. \textbf{c)} Baseline model coefficients. }
    \label{fig:fig10}
\end{figure}





\begin{figure}
    \centering
    \includegraphics[width=\linewidth]{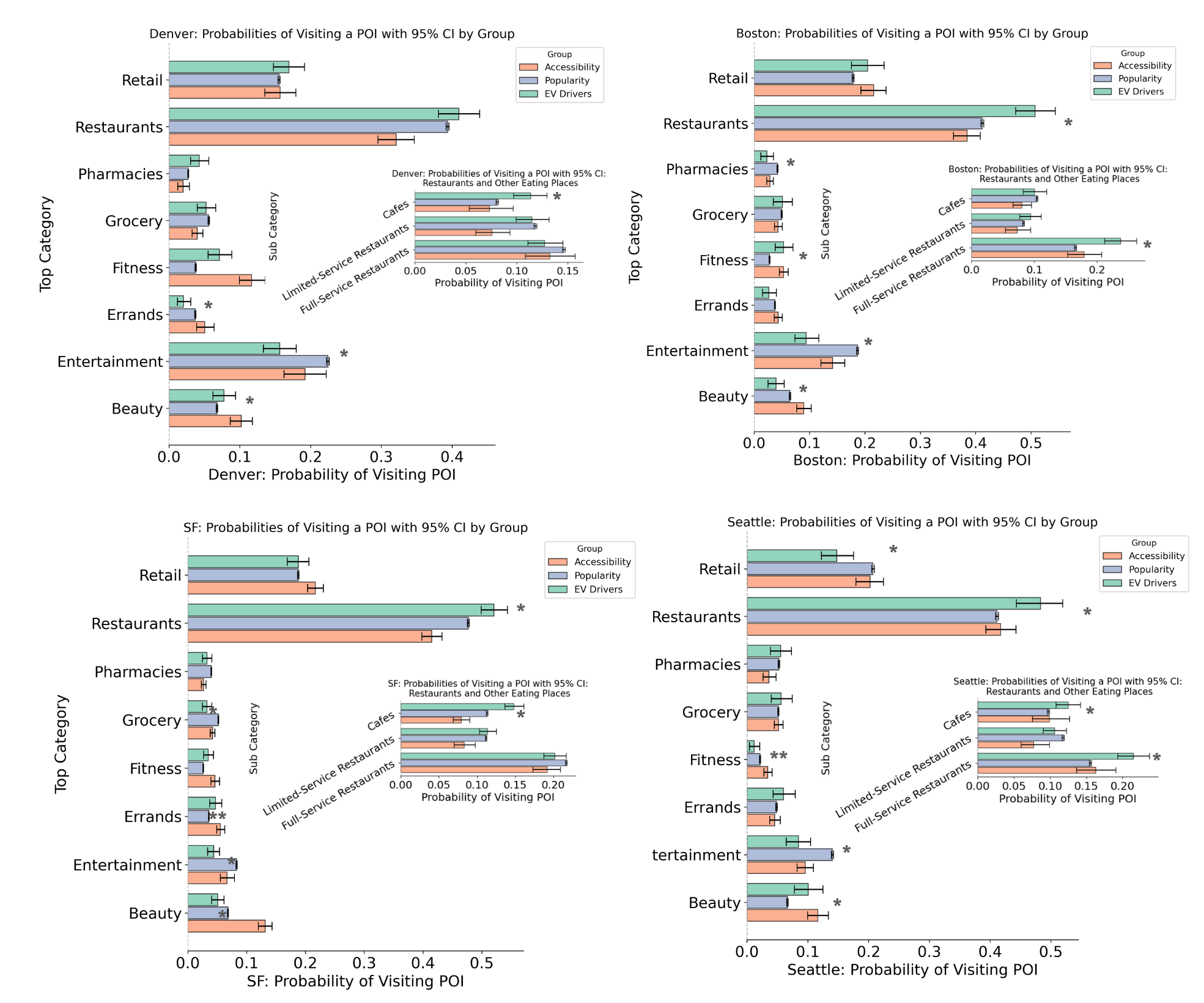}
    \caption{\textbf{Probability of Visiting POI during an EVCS }Probability of visitation by POI type for all four cities, given a POI visit within 250m of an EVCS. Three groups labeled by color:  EV drivers (during a charging session), all users that visit a POI within the EVCS buffer zone (popularity), and the overall prevalence of each POI type (accessibility). EV drivers show significantly higher visitation to restaurants and cafes during charging. Asterisk (\textbf{*}) indicates significant difference between EV driver and study population behavior ($p<0.01$), and (\textbf{**}) indicates ($p<0.05$).}
    \label{fig:fig3_supp}
\end{figure}